# A Machine Learning–Based Framework to Shorten the Questionnaire for Assessing Autism Intervention

Audrey Dong[1†], Claire Xu[2], Samuel R. Guo[3], Kevin Yang[4], Xue-Jun Kong[5,6]

[1] Emma Willard School, Troy, NY 12180
[2] The Harker School, San Jose, CA 95129
[3] Newton South High School, Newton Center, MA 02459
[4] Fairview High School, Boulder, CO 80305
[5] Fetal-Neonatal Neuroimaging and Developmental Science Center, Department of Newborn Medicine, Boston Children's Hospital, Boston, MA 02115
[6] Athinoula A. Martinos Center for Biomedical Imaging, Massachusetts General Hospital, Harvard Medical School, Charlestown, MA 02129

***Abstract*** Caregivers of individuals with autism spectrum disorder (ASD) often find the 77-item Autism Treatment Evaluation Checklist (ATEC) burdensome, limiting its use for routine monitoring. This study introduces a generalizable machine learning framework that seeks to shorten assessments while maintaining evaluative accuracy. Using longitudinal ATEC data from 60 autistic children receiving therapy, we applied feature selection and cross-validation techniques to identify the most predictive items across two assessment goals: longitudinal therapy tracking and point-in-time severity estimation. For progress monitoring, the framework identified 16 items (21% of the original questionnaire) that retained strong correlation with total score change and full subdomain coverage. We also generated smaller subsets (1–7 items) for efficient approximations. For point-in-time severity assessment, our model achieved over 80% classification accuracy using just 13 items (17% of the original set). While demonstrated on ATEC, the methodology—based on subset optimization, model interpretability, and statistical rigor—is broadly applicable to other high-dimensional psychometric tools. The resulting framework could potentially enable more accessible, frequent, and scalable assessments and offer a data-driven approach for AI-supported interventions across neurodevelopmental and psychiatric contexts.

## I. INTRODUCTION

Autism Spectrum Disorder (ASD) is an early onset, lifelong, neurodevelopmental disorder, influenced by a combination of genetic and environmental factors. The early signs might present within the first 12 months, however, a reliable diagnosis by an experienced clinician is expected around 24 months [1]. ASD can be characterized by restricted repetitive behaviors, and a lack of social communication. People with ASD often present several comorbidities, ranging from being nonverbal (25~35% of people with ASD are minimally verbal [2]) or having much difficulty communicating [3], to social skill deficits [4] and often have a lack of cognitive awareness [5].

Some commonly seen comorbidities include ADHD [6], epilepsy [7], and intellectual disability [8], just to name a few.

The effectiveness of therapy for ASD has often been difficult to quantify. Autism Treatment Evaluation Checklist (ATEC) is a commonly used caregiver-reported quantitative tool to track symptom changes with a therapy [9]. It contains 77 questions to cover a broad spectrum of symptoms. Also, to minimize under-coverage bias from each question, such questionnaires were often designed with multiple similar questions to help narrow down results [9]. ATEC consists of 4 subtests with higher scores representing a higher severity: Speech/Language/Communication (0–28), Sociability (0–40), Sensory/Cognitive awareness (0–36), and Health/Physical/Behavior (0–75). The questions were collected from a dozen researchers and clinicians. The preliminary version was validated by parents and professionals before finalizing the design [10].

Current research indicts that caregiver-reported ATEC scores could have a moderate correlation with clinician-rated CARS (Childhood Autism Rating Scale), a well-established questionnaire for diagnosing ASD [11, 12]. ATEC could be useful in screening flags and monitoring response to intervention, alongside clinician-administered assessments (e.g., ADOS-2, CARS-2), clinical interview/DSM-5 criteria, and measures of adaptive functioning (e.g., Vineland), cognition, language, and comorbidities when determining diagnosis and required supports.

Even with all the merits that come with ATEC, some caregivers find it overwhelming and time-consuming to fill out, which can reduce completion rates and introduce unreliable answers such as rushed or missing items. As AI-enabled, cost-efficient interventions become more accessible, validated short-form questionnaires are increasingly important to support frequent assessment alongside clinician-rated measures when clinical decisions are needed.

Our research aims to develop a machine learning-based framework to shorten the longitudinal study-based questionnaire. We focused on the ATEC questionnaire results from 60 autistic children (mean age 5.1 ± 1.3 years at the start of the treatment).

First, we used Matplotlib [13] and Seaborn [14] for exploratory data visualizations. Then we formulated a ML-based framework to shorten the questionnaire using R [15] and multiple linear regression models. We trained models and conducted random shuffle cross-validation. The ML model selected a set of high indicative questions correlating with the effectiveness of the therapy. We then computed the statistical significance (p-value) of each selected question. Systematically, we shortened the ATEC questionnaire to 16 questions while maintaining good question coverage. We further expanded our ML framework's capability to shortlist questions for Point-in-Time diagnosis by selecting questions best reflecting the severity level of the symptoms. We achieved 80+% accuracy by using 13 out of the 77 questions.

The methodological strengths and innovations from the research can be summarized as follows:

- Efficient Longitudinal Assessment and Actionable Insights on the Next Steps of the Intervention: use machine learning-based framework to automatically compute the progress in therapy within seconds with statistical rigor as opposed to the manual time-consuming assessments. Adjust intervention plan based on actionable insights.

[†] Present address: University College London, Gower Street, London WC1E 6BT, UK

- Questionnaire Shortening: robust analysis combining data visualization (Matplotlib and Seaborn) with multiple linear regression models in R. Random shuffle cross-validation was conducted to ensure reliability.

- Tailored Assessment Design: offer a practical solution to customize assessment for diverse user groups undergoing different intervention and therapeutic treatments.

## II. SURVEY OF AUTISM EVALUATION TOOLS

With the advancement of new technologies in AI, robotics and XR, there has been a significant impact on the diagnosis and intervention of autism. The corresponding evaluation questionnaires also need to be evolved to allow ease of operations for more frequent assessment, or even solely use AI to observe the behavior of autistic children [16-18].

We first surveyed commonly used questionnaires for Autism Diagnosis and Intervention Evaluation (see Table 1).

**Table 1.** List of Well-Adopted Questionnaires for Autism Diagnosis and Intervention Evaluation

| Name | Year | Categories/Areas | Number of Items |
|---|---|---|---|
| E-2 [19] | 1971 | ● Pregnancy and Birth Complications<br>● Social skills<br>● Speech development<br>● Atypical Behaviors<br>● Medicines, Diets, and Supplements<br>● Miscellaneous Therapy and Conditions | 109 |
| Childhood Autism Rating Scale (CARS-2) [20] | 1980 | ● Social Relationships<br>● Rigidness in behaviors<br>● Communication<br>● Activity level<br>● Level and consistency of intellectual responses and Social Responses | 15 |
| Autism Behavior Checklist (ABC) [21] | 1980 | ● Sensory Behaviors (9 Questions)<br>● Relating Behaviors (12 Questions)<br>● Body and Object Use Behaviors (12 Questions) | 33 |
| Autism diagnostic observation schedule (ADOS) [22] | 1989 | ● A standardized protocol for observation of social and communicative behavior associated with autism | 29 |
| Autism Diagnostic Interview (ADI) (revised ADI-R) [23] | 1989 | ● Communication<br>● Sociability<br>● Imagination<br>● Lack of Shared Enjoyment | 93 |
| Checklist for Autism in Toddlers (CHAT) [24] | 1992 | ● Play Behavior<br>● Understands Pointing<br>● Sociability<br>● Motor development<br>● Eye contact | 14 |
| Modified Checklist for Toddlers (M-CHAT) [25] | 2001 | ● Understands Nonverbal Cues<br>● Understands Verbal Cues<br>● Understands Social Norms<br>● Stimulated/Overstimulated by Sensory Activities | 20 |
| Gilliam Autism Rating Scale (GARS) [26] | | ● Stereotyped Behavior<br>● Communication Behavior<br>● Social Interaction Behavior<br>● Play Behavior | 42 |
| Autism Screening Instrument for Educational Planning (ASIEP) [27] | 1999 | ● Autism Behavior Checklist<br>● Sample of Vocal behavior<br>● Interaction Assessment<br>● Education Assessment<br>● Prognosis of Learning Rate | 5 areas, each with multiple questions, observations, or evaluation items |
| Autism Treatment Evaluation Checklist (ATEC) [9]. | 1999 | ● Speech/Language/Communication<br>● Sociability<br>● Sensory Cognitive Awareness<br>● Health/Physical/Behaviors | 77 |
| Asperger's Syndrome Diagnostic Interview (ASDI) [28] | 2001 | ● Social interaction<br>● Obsessive interests<br>● Rigid Behaviors<br>● Communication Problems<br>● Motor Clumsiness | 20 |

We decided to use ATEC to develop a robust ML-based framework to shorten the questionnaire. During the framework design process, we also kept in mind the other questionnaires so that the framework can be applicable to other questionnaires for longitudinal study, such as assessing the effectiveness of treatment or progression of symptoms.

## III. PROPOSED ML-BASED FRAMEWORK

In order to design a robust ML framework that supports the following two goals, we started with questions interpretation and grouping (Section A in Fig. 1) and exploratory data visualizations (Section B in Fig. 1)

- Correlate with the longitudinal changes (illustrated in Section C.1 in Fig.1 and text below)
- Point-in-time diagnosis (illustrated in Section C.2 in Fig. 1 and text below)

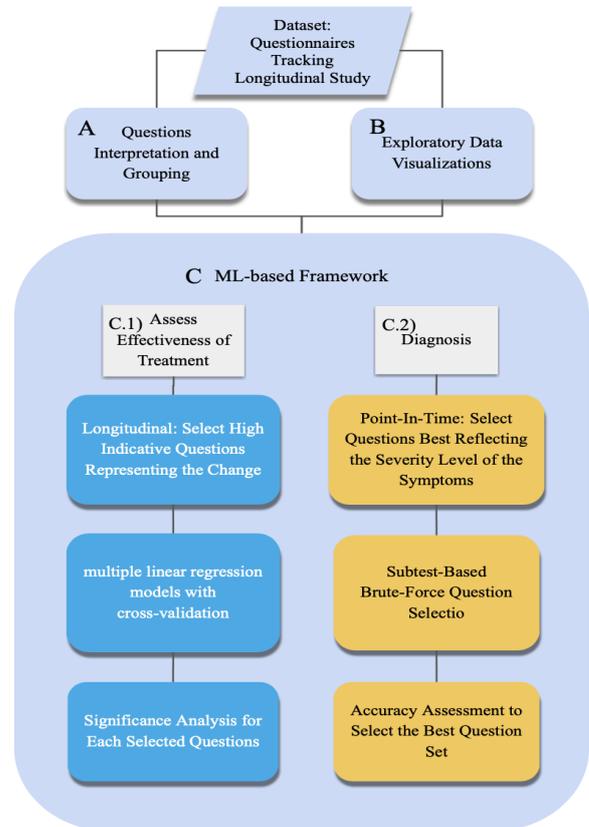

**Fig. 1.** ML-Based framework to shorten the questionnaire used for longitudinal study.

## A. Questions Interpretation and Grouping

Within each ATEC subtest, we first grouped the questions based on their similarities/connections into mini-clusters. The grouping is based on the assessed skills or conditions. Most of these mini-clusters don't overlap, with the notable exception of dietary problems and gastrointestinal problems as both can be related or separate issues. Please refer to table 3 below for details.

Our hypothesis is that our shortlisted selections should fairly cover the various mini-clusters of questions, and also correlate with the areas that therapy focuses on. For the areas that are outside of the therapy focus, we may not see questions on our shortlist.

## B. Exploratory Data Visualizations

We collected data from Star Bridge Organization (zhmxxq.com), specializing in Occupational Therapy, Behavioral Therapy, Speech Language Therapy, and Social Thinking. The main therapy models used are LSP (Learning style profile for Children with Autism Spectrum Disorder) [29] and a modified version of SCERTS (Social Communication, Emotional Regulation, Transactional Support).

ATEC responses were collected from a total of 60 children diagnosed with ASD who had undergone intervention. Each subject took two assessments: initial assessment at the onset of their treatment (mean age $5.1 \pm 1.3$ years), then a subsequent assessment based on which their therapy methods (LSP vs SCERTS) were adjusted according to the result (mean age $5.9 \pm 1.4$ years). The mean duration between the two assessments is $36 \pm 17$ weeks.

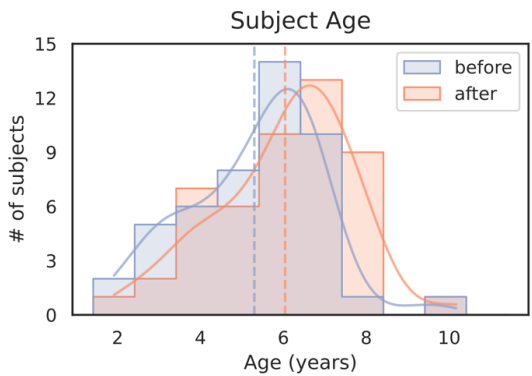

**Fig. 2.** The mean age is $5.1 \pm 1.3$ years before and $5.9 \pm 1.4$ years after. The distribution is skewed to the left (younger), peaking between 6 and 7 years both before and after the treatment.

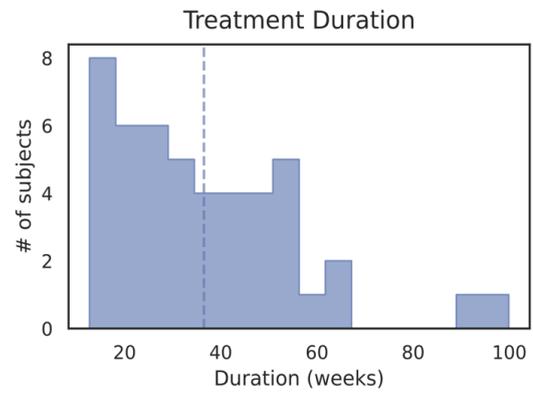

**Fig. 3.** The mean duration of the treatment is $36 \pm 17$ weeks. The distribution is strongly skewed to the right (longer) and ranges from 16 to 65 weeks, with two outliers (92 and 99 weeks, respectively).

In this paper, we calculated the score change by subtracting the score after from the score before ($\Delta$ = before – after). In the ATEC, a higher score indicates higher symptom severity. Therefore, with this method of calculating the change in scores, a positive change indicates an improvement in symptoms, while a negative change indicates deterioration. We calculated the average improvement percentage per subject (improved score/before treatment score). The subtests were ranked from most to least improved: Speech/Language/Communication ($40.6 \pm 27.1\%$), Sensory/Cognitive awareness ($30.1 \pm 43.0\%$), Sociability ($17.6 \pm 70.6\%$), and Health/Physical/Behavior ($-11.7 \pm 244\%$).

This analysis result aligns with the characteristics of the LSP educational intervention and suggests the adjustment in the next step of the intervention. LSP focuses on cultivating professionals and adults to empower children in autonomously applying social, communication, and interactional skills, without relying on external prompt from adults and peers. The data analysis showed significant improvement in speech, language and communication. Since social disorder presents significant challenge for children with ASD, substantial improvement in this area is expected after longer term intervention. Health/Physical/Behavior didn't show improvement and, in some cases, even worsened. This suggests the possible consideration in the next step on dietary changes or medication under the guidance of a doctor.

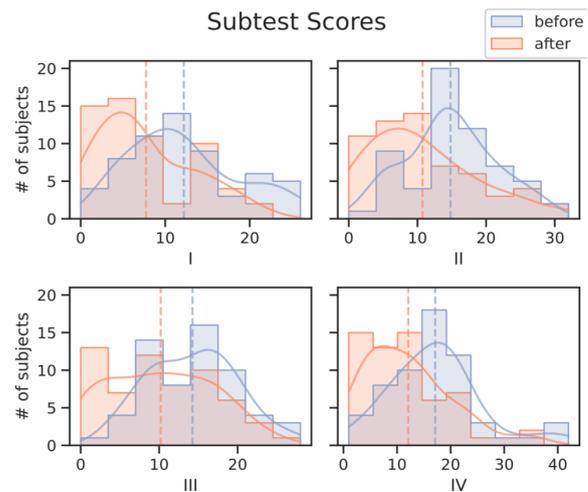

**Fig. 4.** The before and after treatment scores for subscale I (0–28), II (0–40), III (0–36), and IV (0–75). The mean subscale scores for subtests I, II, III, and IV shifted, respectively, from $12.2 \pm 6.5$ to $7.7 \pm 5.6$, from $14.8 \pm 6.6$ to $10.8 \pm 8.0$, from $14.3 \pm 5.7$ to $10.2 \pm 6.7$, and from $17.1 \pm 8.6$ to $12.1 \pm 8.2$ during treatment, with mean per subject improvements of $4.5 \pm 3.7$, $4.1 \pm 8.2$, $4.0 \pm 5.4$, and $5.0 \pm 6.4$, respectively, all statistically significant ($p < 0.0002$).

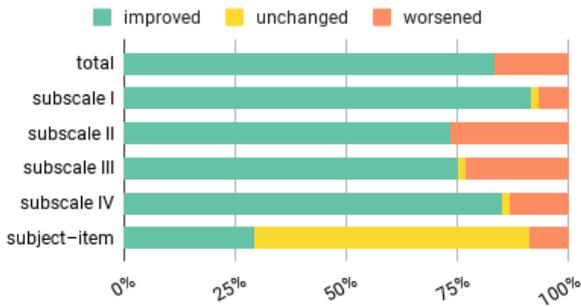

**Fig. 5**. Overall (as measured by the total ATEC score), 50 subjects (83%) improved while 10 subjects (17%) worsened. Out of all the 60 subjects' item scores for the 77 individual items (totaling 4620 subject–item entries), 29.5% showed improvement, 61.5% remained unchanged, and 8.98% showed deterioration.

The relationships between change in total score and changes in item scores are generally positive, which implies that improvements in item scores generally correlate with improvements in the total score, and vice versa.

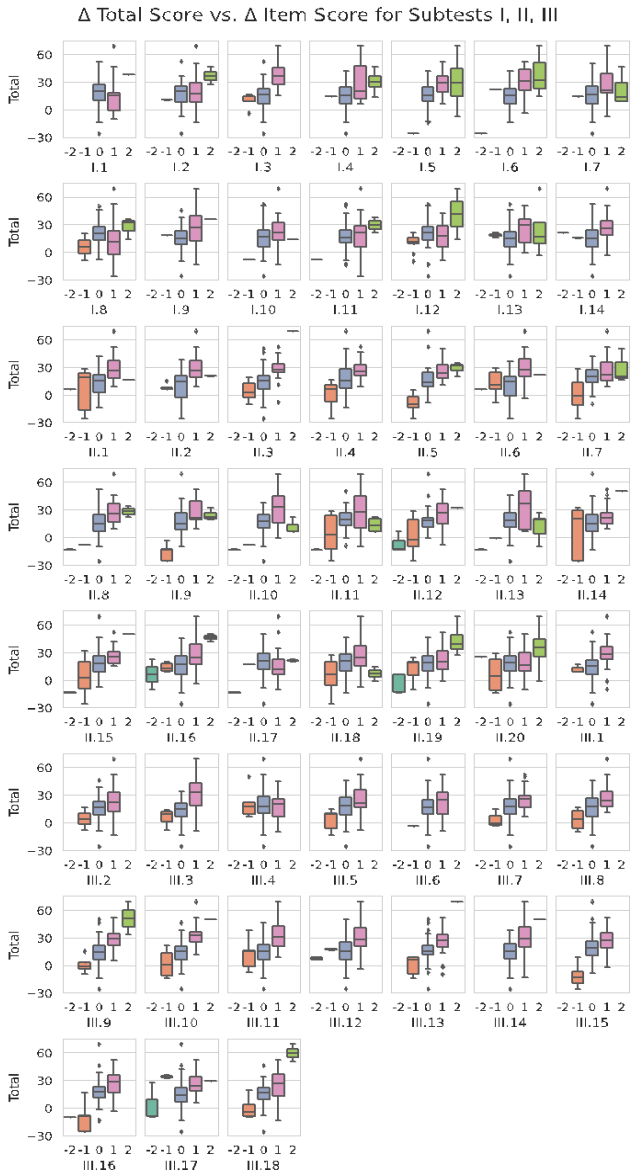

**Fig. 6**. Relationships between the changes in total score and item scores for subtests I, II, and III.

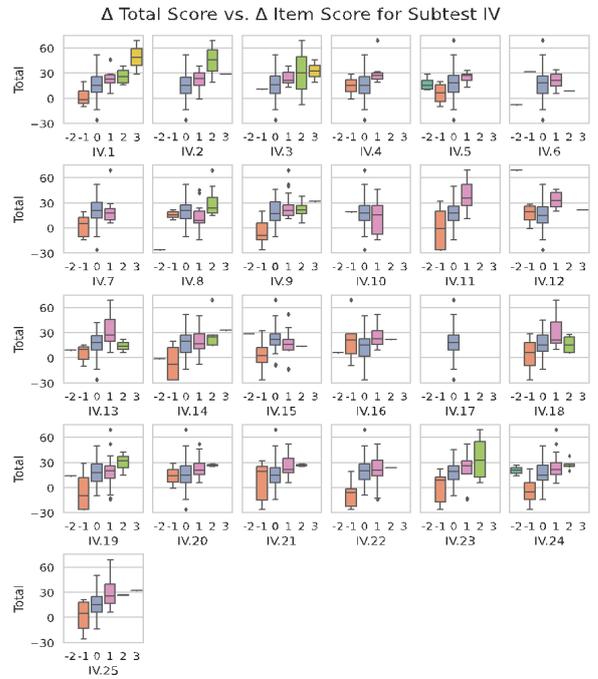

**Fig. 7**. Relationships between the changes in total score and item scores for subtest IV.

*C. Machine Learning Based Question Selection Framework*

In this study, we developed two question selection approaches to achieve the respective goals.

Goal 1): to assess the longitudinal change (i.e., the effectiveness of the Autism therapy), the framework selects high indicative questions representing the change over time.

Goal 2): to facilitate the Point-In-Time diagnosis, the framework selects questions best reflecting the severity level of the symptoms.

*1) Shortening Questionnaire for Longitudinal Study*

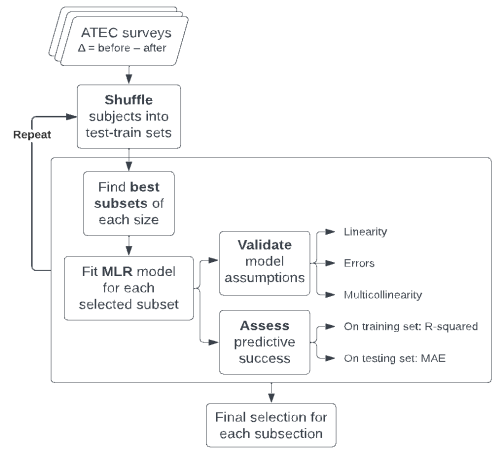

**Fig. 8**. Flow chart of the question selection and evaluation process.

We correlated the changes in item scores to the total change to find items that would best represent the total treatment effect. To avoid overfitting and provide variance so that the results remain general, we randomly shuffled the subjects to break any patterns in the data that might affect the performance of the models during cross-validation. Then the data was split to create multiple 70%–30% test–train splits [30]. For each shuffle, out of the total 77 questions, we selected subsets of 1–7 question items that would best predict the change in total score compared to other subsets of the same size. We fitted and assessed multiple linear regression models with the selected subsets. Finally, we summarize our findings and recommend shortened versions of the questionnaire.

The multiple linear regression (MLR) is a standard and widely used statistical technique that models the relationship between a quantitative response variable and multiple quantitative predictor variables by fitting a linear equation to the data [31]. The regression line is defined to be $\mu_y = \beta_0 + \beta_1 x_1 + \beta_2 x_2 + \cdots + \beta_p x_p$ for each predictor variable $x_i$ ($1 \leq i \leq p$), where each parameter $\beta_i$ ($0 \leq i \leq p$) is estimated using the least squares technique, which minimizes the residual sum of squares (RSS). In each of our MLR models, the predictor variables are the changes in the scores of the question items in the selected subset, and the response variable is the change in total score of the ATEC response.

MLR is the most suitable technique for our purpose as it mirrors the original scoring structure of the ATEC. The total score of as given ATEC response is simply the aggregate of the scores of survey's individual question items. In our models, to estimate the change in total score, only the score changes of the selected question items will be considered, but each of them will be scaled by a coefficient $\beta_i$ ($1 \leq i \leq p$) before being aggregated. We will compute the $R^2$ (coefficient of determination) [32] and mean absolute error (MAE) [33] for each MLR model we present to measure its predictive success. Moreover, because the technique involves few parameters, the results are easy to interpret and less prone to overfit the data.

We used the best subsets algorithm from the R package leaps [34] to select subsets of questions from which to estimate the total score. The algorithm finds the subsets of questions that generate the least RSS compared to subsets of the same size by performing exhaustive search on all possible subsets of the given sizes. We then fitted models from these selected subsets using MLR and assessed their performance.

All the assumptions of MLR should be reasonably met by our dataset [31]. The linearity assumption, which requires that the relationship between the response variable and each predictor variable is linear and that there are no severe outliers, is supported for most items by the box plots in Figs 6–7, especially for the items selected in the best subsets. The error assumptions of a specific model will be checked below as an example. The risk of multicollinearity (high correlation between the predictor variables), checked numerically below, should be very low because the number of items chosen in each subset as predictor variables (1–7) is small compared to the total number of items in the ATEC (77) so that strongly correlated items should not be selected together by the best subsets algorithm as they should not increase the predictive strength of the model.

**Table 2**. The subsets of questions selected for each shuffle and the statistics of the MLR models fitted

| Shuffle | Size | Items | Training Set | | Testing Set |
| --- | --- | --- | --- | --- | --- |
| | | | Multiple $R^2$ | Adjusted $R^2$ | MAE |
| 1 | 1 | 11.90 + 15.11 **I.6** | 0.47 | 0.46 | 12.1 |
| | 2 | 10.36 + 14.14 **I.6** + 10.64 **II.8** | 0.70 | 0.68 | 12.9 |
| | 3 | 8.27 + 11.74 **I.6** + 11.44 **II.8** + 10.61 **III.18** | 0.81 | 0.79 | 12.4 |
| | 4 | 8.40 + 10.97 **I.6** + 9.43 **II.4** + 9.42 **II.8** + 9.42 **III.18** | 0.88 | 0.87 | 11.3 |
| | 5 | 7.01 + 10.20 **I.6** + 6.19 **II.1** + 8.49 **II.8** + 6.49 **III.5** + 9.87 **III.8** | 0.92 | 0.90 | 13.4 |
| | 6 | 6.62 + 10.31 **I.6** + 8.36 **II.4** + 8.87 **II.8** + 6.07 **III.5** + 6.32 **III.8** + 7.82 **III.18** | 0.95 | 0.94 | 10.1 |
| | 7 | 4.74 + 9.05 **I.1** + 9.62 **I.6** + 9.18 **II.4** + 8.92 **II.8** + 4.71 **III.5** + 8.88 **III.8** + 8.34 **III.18** | 0.96 | 0.95 | 10.0 |
| 2 | 1 | 13.16 + 15.96 **III.9** | 0.40 | 0.39 | 13.8 |
| | 2 | 11.63 + 13.19 **III.9** + 16.68 **III.15** | 0.64 | 0.65 | 11.2 |
| | 3 | 12.79 + 11.67 **II.7** + 10.19 **III.9** + 14.69 **IV.11** | 0.80 | 0.79 | 15.0 |
| | 4 | 8.59 + 7.82 **II.5** + 8.27 **II.12** + 13.98 **III.11** + 10.80 **IV.23** | 0.87 | 0.86 | 11.8 |
| | 5 | 8.20 + 7.79 **II.5** + 7.80 **II.12** + 2.08 **II.20** + 13.47 **III.11** + 10.74 **IV.23** | 0.90 | 0.89 | 11.7 |
| | 6 | 5.42 + 10.33 **II.1** + 7.27 **II.11** + 12.45 **III.8** + 6.72 **III.13** + 10.24 **III.16** + 9.75 **IV.3** | 0.93 | 0.93 | 7.4 |
| | 7 | 5.20 + 9.77 **II.1** + 4.09 **II.3** + 7.48 **II.11** + 12.63 **III.8** + 5.22 **III.13** + 8.86 **III.16** + 8.79 **IV.3** | 0.95 | 0.94 | 6.4 |
| 3 | 1 | 11.54 + 13.69 **II.5** | 0.44 | 0.42 | 11.9 |
| | 2 | 11.72 + 12.05 **I.6** + 15.67 **III.15** | 0.61 | 0.59 | 12.8 |
| | 3 | 10.08 + 11.60 **I.6** + 9.51 **II.8** + 13.82 **III.15** | 0.74 | 0.72 | 12.6 |
| | 4 | 10.14 + 10.97 **I.6** + 6.96 **II.4** + 8.68 **II.8** + 11.27 **III.15** | 0.81 | 0.79 | 12.5 |
| | 5 | 9.10 + 10.29 **I.6** + 6.62 **II.4** + 8.94 **II.8** + 10.23 **III.15** + 6.34 **III.18** | 0.86 | 0.84 | 10.6 |
| | 6 | 7.69 + 9.09 **I.6** + 6.51 **II.4** + 8.89 **II.8** + 10.92 **III.15** + 7.72 **III.18** + 4.45 **IV.3** | 0.89 | 0.88 | 10.1 |
| | 7 | 3.78 + 9.56 **II.1** + 7.60 **II.3** + 6.99 **II.11** + 8.35 **III.8** + 10.75 **III.16** + 8.21 **IV.3** + 3.54 **IV.9** | 0.92 | 0.90 | 5.0 |
| 4 | 1 | 12.70 + 20.08 **III.10** | 0.49 | 0.48 | 12.3 |
| | 2 | 12.21 + 8.67 **II.12** + 16.57 **III.10** | 0.62 | 0.60 | 10.8 |
| | 3 | 7.45 + 12.94 **II.9** + 9.62 **II.19** + 12.75 **III.18** | 0.77 | 0.77 | 12.5 |
| | 4 | 6.38 + 12.04 **II.9** + 7.55 **II.19** + 8.02 **III.9** + 10.91 **III.18** | 0.84 | 0.83 | 11.2 |
| | 5 | 5.26 + 8.90 **II.9** + 6.89 **II.15** + 7.86 **II.20** + 12.11 **III.9** + 9.75 **III.18** | 0.90 | 0.89 | 9.4 |
| | 6 | 4.96 + 8.65 **I.6** + 7.57 **II.4** + 7.40 **II.10** + 5.32 **II.19** + 9.19 **III.18** + 5.41 **IV.9** | 0.93 | 0.92 | 7.8 |
| | 7 | 4.57 + 8.16 **I.6** + 8.00 **II.4** + 5.79 **II.10** + 5.00 **II.19** + 4.21 **III.5** + 8.91 **III.18** + 5.20 **IV.9** | 0.95 | 0.94 | 7.8 |
| 5 | 1 | 11.80 + 14.96 **II.5** | 0.44 | 0.43 | 11.2 |
| | 2 | 11.01 + 14.16 **I.6** + 12.25 **II.5** | 0.66 | 0.64 | 13.1 |

| | 3 | 13.38 + 14.45 **I.6** + 9.03 **II.8** + 9.88 **III.13** | 0.80 | 0.79 | 15.7 |
| | 4 | 13.62 + 12.71 **I.6** + 6.88 **II.6** + 7.01 **II.12** + 9.24 **III.10** | 0.87 | 0.86 | 15.6 |
| | 5 | 12.39 + 10.09 **I.6** + 7.04 **II.6** + 7.63 **II.12** + 8.84 **III.10** + 5.47 **IV.21** | 0.91 | 0.90 | 14.0 |
| | 6 | 10.32 + 12.75 **I.6** + 6.40 **II.6** + 5.83 **II.10** + 3.51 **II.19** + 7.97 **III.16** + 3.83 **IV.9** | 0.95 | 0.94 | 12.8 |
| | 7 | 10.05 + 10.59 **I.6** + 5.46 **II.2** + 7.09 **II.6** + 3.19 **II.11** + 5.27 **II.12** + 8.01 **III.10** + 4.05 **IV.18** | 0.96 | 0.95 | 12.8 |
| | 1 | 14.70 + 18.18 **III.10** | 0.43 | 0.41 | 11.2 |
| | 2 | 12.51 + 12.35 **II.15** + 16.16 **III.9** | 0.65 | 0.63 | 13.4 |
| | 3 | 8.75 + 12.96 **III.10** + 12.12 **III.18** + 10.83 **IV.25** | 0.79 | 0.77 | 13.7 |
| 6 | 4 | 8.92 + 7.18 **II.15** + 10.42 **III.10** + 10.91 **III.18** + 10.41 **IV.25** | 0.86 | 0.84 | 11.4 |
| | 5 | 6.84 + 8.55 **II.15** + 5.74 **II.20** + 10.38 **III.9** + 9.40 **III.18** + 8.68 **IV.25** | 0.90 | 0.88 | 11.8 |
| | 6 | 5.78 + 6.61 **I.6** + 8.59 **II.2** + 5.39 **II.11** + 8.40 **II.15** + 11.36 **III.9** + 6.51 **IV.23** | 0.93 | 0.92 | 10.5 |
| | 7 | 6.66 + 10.79 **II.9** + 4.41 **II.19** + 12.37 **III.10** + 7.42 **III.18** + -6.74 **IV.12** + 7.66 **IV.22** + 6.19 **IV.25** | 0.95 | 0.94 | 9.1 |

Table 2 summarizes the performance of multiple linear regression (MLR) models fitted on different random shuffles of the dataset, each using 1–7 selected ATEC items to predict total score change. For each shuffle, the table reports training fit (Multiple R² and Adjusted R²) and testing accuracy (MAE), illustrating how reduced item sets capture treatment effects while maintaining generalizability.

**Table 3**. The statistics results of the example MLR model (shuffle 1, subset size 5)

| Coefficient | Estimate | Std. Error | t value | P-value |
|---|---|---|---|---|
| (Intercept) | 7.831 | 1.127 | 6.95 | 3.82e-08 |
| I.6 | 11.404 | 1.471 | 7.755 | 3.45e-09 |
| II.1 | 6.68 | 1.486 | 4.496 | 6.93e-05 |
| II.8 | 7.295 | 1.444 | 5.051 | 1.28e-05 |
| III.5 | 9.588 | 1.674 | 5.727 | 1.61e-06 |
| III.8 | 14.99 | 1.899 | 7.894 | 2.29e-09 |

Residual standard error: 6.317 on 36 degrees of freedom; Multiple R-squared: 0.9153; Adjusted R-squared: 0.9035; F-statistic: 77.8 on 5 and 36 DF, p-value: < 2.2e-16.

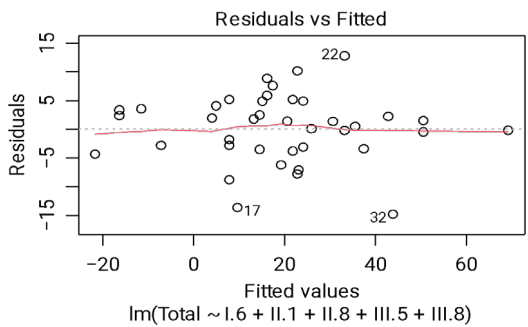

**Fig. 9**. Residual plot of the example model. The errors are independent (residuals are scattered without pattern above and below the zero line), have a mean of zero (residuals are centered around zero), and have a constant standard deviation (residuals show constant spread above and below the zero line from left to right).

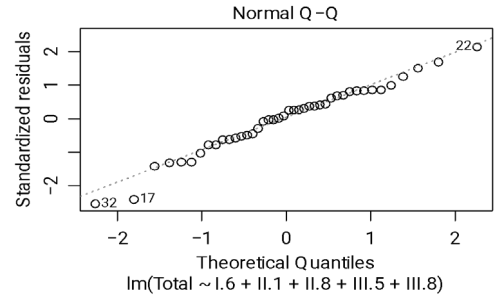

**Fig. 10**. Normal Q-Q plot of the example model's residuals. The residuals are all close to the identity line except two outliers at the bottom left, suggesting that the errors are roughly normally distributed.

We now consider as an example the MLR model fitted from the first shuffle with 5 predictor variables. The total score change of a given ATEC response $i$ is modeled to be $7.831 + 11.404(I.6) + 6.680(II.1) + 7.295(II.8) + 9.588(III.5) + 14.990(III.8) + \varepsilon_i$, where the item numbers indicate the score change of that item and where the errors are assumed to be independent, have mean zero and constant standard deviation, and be normally distributed. All the assumptions for MLR are reasonably met according to the diagnostics: the linearity assumption is reasonably satisfied for the predictor variables chosen (Fig. 6-7) and the error assumptions are validated by the residual plot (Fig. 9) and the residual normal Q-Q plot (Fig. 10). The coefficient $\hat{\beta}_1 = b_1 = 11.404$, for example, means that for each unit increase in the score change of item I.6, the total score change increases by 11.404 units on average, holding fixed the score changes of all the other items.

The data provides compelling evidence to generalize the relationship stated to our population (Table 3). The p-value of the F-statistic is less than $2.2 \times 10^{-16}$, meaning that at least one of the predictor variables $x_i$ $(1 \leq i \leq p)$ is important for predicting the total score. The t-test for slope was performed on each coefficient $\beta_i$ $(1 \leq i \leq p)$. Here we consider the term with item I.6 as an example. The null hypothesis $H_0$ states that there is no true linear relationship between the score change of item I.6 and the total score change ($\beta_1 = 0$), and the alternative hypothesis $H_A$ states that a true linear relationship exists between them ($\beta_1 \neq 0$). The standardized score of the test statistics is $t = \frac{b_1 - 0}{SE} = \frac{11.404 - 0}{1.471} = 7.755$, and the p-value of the t-statistic is $3.45 \times 10^{-9}$. Since the p-value is less than 0.05, we reject the null hypothesis, and we have sufficient evidence that the I.5 term is significant in the presence of all the other predictor variables. Further, since the p-values of all $\beta_i$ $(1 \leq i \leq p)$ are less than 0.05, we have sufficient evidence that each of the predictor variables is significant in the presence of all the other predictor variables in the model. The R² (coefficient of determination), which measures the proportion of the variation in the response that is predictable from the predictor variables [32], is 0.9153, and the MAE [33], calculated in the testing dataset, is 13.35%. The variance inflation factor (VIF) measures collinearity in multiple regressions by regressing a predictor on all the other predictors in the model [35]. The VIFs for all five selected items are between 1 and 1.3, suggesting low multicollinearity. For all the MLR models, the p-values for all coefficients are statistically significant (<0.01), and the VIFs for all selected items are under 1.6, indicating that multicollinearity is not a concern.

**Table 4**. Final item selections for question mini-clusters under each ATEC subtest. Items that appeared in selected subsets in any shuffle have their p-values from matched pairs t-tests for treatment effect displayed. The final selection is based on the monotonically related effect size and p-value

| ATEC subtests | Questions Mini-Cluster | Item # | Shuffle 1 | Shuffle 2 | Shuffle 3 | Shuffle 4 | Shuffle 5 | Shuffle 6 | p | Final Selection |
|---|---|---|---|---|---|---|---|---|---|---|
| I. Speech/ Language/ Communication | Basic | 1, 2, 3 | 1 | | | | | | 1: 0.00049 | 1 |
| | Intermediate | 9, 10, 11 | | | | | | | | |
| | Advanced | 12, 13, 14 | | | | | | | | |
| | Multiword | 4, 5, 6, 7, 8, 10, 12 | 6 | | 6 | 6 | 6 | 6 | 6: 0.010 | 6 |
| II. Sociability | World of their own | 1,2,3,9 | | 1, 3 | 1, 3 | | 2, 6 | 2, 6 | 1: 0.029 2: 0.000020 3: 0.0027 6: 0.077 | 2 |
| | Uncooperative | 4,14,15, 10 | 4 | | 4 | 4, 10, 15 | 10 | 15 | 4: 0.12 10: 0.0027 15: 0.63 | 10 |
| | Social norms | 5, 8,12,13,17 | 8 | 5, 12 | 5, 8 | 12 | 5 8, 12 | 12 | 5: 0.00049 8: 0.018 12: 0.46 | 5 |
| | Insensitive | 7,9,11,18,19,20 | | 7, 11, 20 | 11 | 9, 19, 20 | 11 | 9, 11, 19, 20 | 7: 0.18 9: 0.0024 11: 0.23 19: 0.25 20: 0.018 | 9 |
| | Lack friends | 16 | | | | | | | | |
| III. Sensory/Cognitive Awareness | Responses | 1,2 | | | | | | | | |
| | Looking | 3,4,18 | 18 | | 18 | 18 | | 18 | 18: 0.0019 | 18 |
| | Appropriate Play | 5,6,7,8 | 5, 8 | 8 | 8 | 5 | | | 5: 0.015 8: 0.0082 | 8 |
| | Awareness | 9,10,11,17 | | 9, 11 | | 9, 10 | 10 | 9, 10 | 9: 0.0011 10: 0.011 11: 0.015 | 9 |
| | Curiosity | 12,13,14,15,16 | | 15, 13, 16 | 15, 16 | | 13, 16 | | 13: 0.0079 15: 0.13 16: 0.033 | 13 |
| IV. Health/ Physical/ Behavior | Bladder Control | 1,2 | | | | | | | | |
| | Gastrointestinal | 3, 4, 5, 7, 8 | | 3 | 3 | | | | 3: 0.0051 | 3 |
| | Energy | 6,9,10 | | | 9 | 9 | 9 | | 9: 0.00033 | 9 |
| | Destructive | 11,12,13,20 | | 11 | | | | 12 | 11: 1 12: 0.54 | |
| | Agitation | 9, 14, 15, 20, 22 | | | | | | 22 | 22: 0.0027 | 22 |
| | Obsession | 18, 24 | | | | | 18 | | 18: 0.0056 | 18 |
| | Rigidness | 19,21,25 | | | | | 21 | 25 | 21: 0.0079 25: 0.018 | 21 |
| | Unhappy/Crying | 16 | | | | | | | | |
| | Epilepsy | 17 | | | | | | | | |
| | Pain sensitivity | 23 | | 23 | | | 23 | | 23: 0.033 | 23 |
| | Sensory Overstimulation | 14,17 | | | | | | | | |

We recorded the appearance of each mini-cluster's items in each shuffle's chosen best subsets (sizes 1–7) in Table 4. The final selection for each mini-cluster is the item with the highest effect size that reaches statistical significance, which translates to the lowest p-value (Table 4) as the Cohen's *d* for two paired samples, $d = \frac{\bar{X_1} - \bar{X_2}}{SD} = \frac{t}{\sqrt{N}}$[36], is monotonically related to the t-value and therefore the p-value.

*2) Shortening Questionnaire for Diagnosing Symptom Severity*

Current research suggests that ATEC could have a strong correlation with CARS (Childhood Autism Checklist), a well-established questionnaire for diagnosing ASD [11], meaning that ATEC could be used as a tool to help practitioners decide whether to give an official diagnosis for autism and what degree of support is required.

An experiment conducted by Geier and Kern with a sample size of 56 children located and diagnosed in the State of Texas who were diagnosed with ASD, had the parentally completed ATEC and the health-care-professional completed CARS survey, and Spearman's Rank Correlation Coefficient was p=0.71 revealing a significant correlation between ATEC and CARS scales.[11]. A Sudanese study found that among a sample of 40 children diagnosed with autism the Spearman's rank correlation coefficient between ATEC and CARS was 0.015 [12]. Therefore, we expanded our ML-based framework to shortlist the questionnaire for screening/diagnosis purposes.

Data preprocessing was conducted according to the following rules:

● Leap Day of a non-leap year: Replaced with March 1 of that year.

● Missing the day of the month but having the year and month: The first day of the month was imputed.

● Subjects with completely missing dates (birthdays or ATEC completion dates) were excluded from the analysis.

Then, we defined the accuracy function based on three severity levels (Mild, Moderate, Severe). Mild was defined to be the range [0 - 39 (2/9 ATEC max score)], moderate was [40 - 89 (1/2 the max score)], and severe was [90 - 180 (max score)]. The total score of the shortlisted questions was proportionally scaled up and evaluated to one of the three severity levels. When the severity level matched the level from the original ATEC result, this shortlist was qualified as accurate.

After preprocessing the data and defining the accuracy function, we then conducted 5 iterations of model tuning.

Iteration 1-4 involved selecting sets of 5 representative questions across the four ATEC subtests: one question from each subtest, along with an additional supplementary question from any subtest. A brute-forced search was conducted over all 9,198,000 (14 questions in subtest I * 20 in subtest II * 18 in subtest III * 25 in subtest IV * (77 total questions – 4 chosen ones)) combinations of questions.

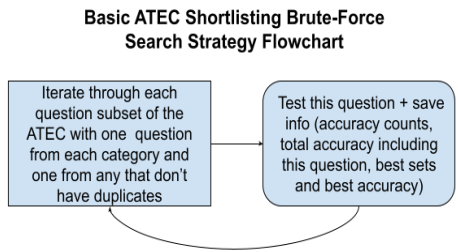

**Fig. 11**. Subtest-Based Brute-Force Question Selection Flowchart

The best question sets (i.e., qualified question sets) were defined as those that reach the same severity conclusion as the original ATEC results (mild, moderate, severe). We kept track of the # of times (i.e., frequency) that each question shown up in all the best question sets. The high frequent questions were chosen to be part of the shortlist (maybe more than 5 questions).

Iteration 1: without age group cohort

Iteration 2: with only samples in 2-5 age group

Iteration 3: with only samples in 6-10 age group

Iteration 4: with both age groups, evenly weighted

Iteration 5: totally different design from iteration 1-4. It randomly sampled any of the 77 questions without needing to select sets of 5 questions, i.e., 1 from each subtest and 1 supplementary.

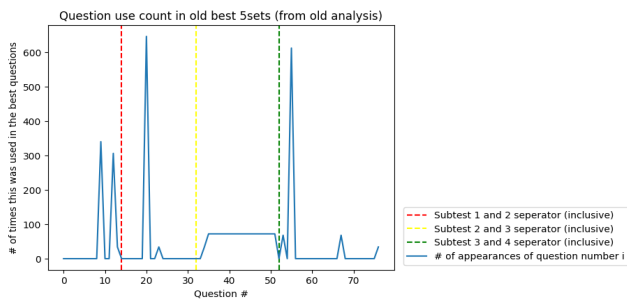

**Fig. 12**: Iteration 1 shortlisted 42 questions that exhibit high frequency across all qualified question sets (y-axis): #9 (I.10), #12 (I.13), #20 (II.7), # 35-51 (III.2 - III.18), and #55 (IV.4). The question sequencing number across 4 ATEC subtests (x-axis) is offset by 1 due to the index starting at 0.

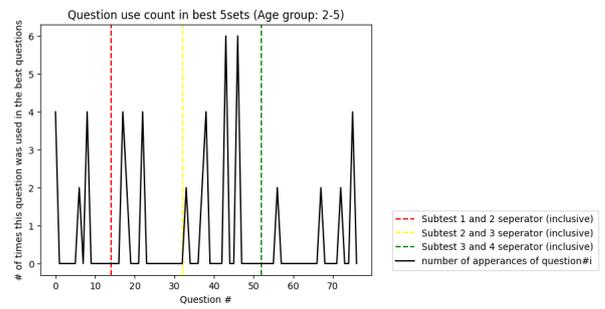

**Fig. 13**. Iteration 2 with only 2-5 age group shortlisted 8 questions that exhibit high frequency across all qualified question sets (y-axis): I.1, I.9, II.4, II.9, III.5, III.10, III.13, and IV.24.

The next analysis is the 6-10 age group and of course, the weighting was still there but the 2-5 age group is absent.

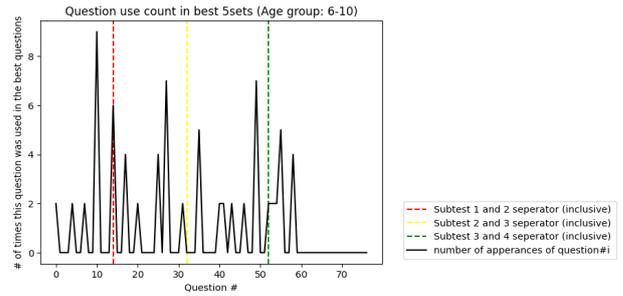

**Fig. 14**. Iteration 3 with only 6-10 age group shortlisted 5 questions that exhibit high frequency across all qualified question sets (y-axis): I.11, II.14, III.16, II.1, and IV.4.

In iteration 4, we considered weighted age groups, making 2-5 and 6-10 age groups each contributing 50% of the accuracy evaluation. If there were A subjects in the 2-5 age group, then each subject in group would contribute $(1/A) \times 50\%$ to the accuracy, and if there were B subjects in the 6-10 age group, then each subject in this age group would contribute $(1/B) \times 50\%$ to the accuracy.

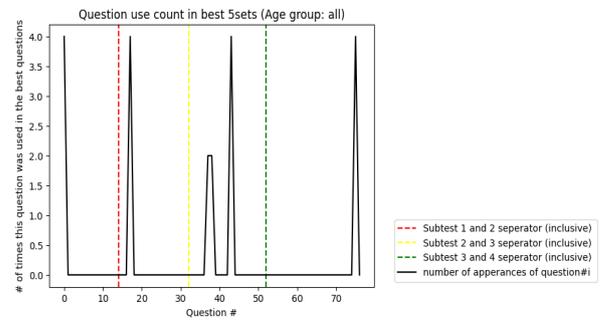

**Fig. 15**. Iteration 4 shortlisted 6 questions that exhibit high frequency across all qualified question sets (y-axis): #0 (I.1), #17 (II.4), #37-38 (III.3, III.4), #43 (III.9), and #75 (IV.24). The question sequencing number across 4 ATEC subtests (x-axis) is offset by 1 due to the index starting at 0.

For iteration 5, since testing all question subsets ($2^{77} = 151,115,727,451,828,647,838,272 \approx 1.511 \times 10^{23}$) would be prohibitively expensive, our experiments tested 1250 random sets of each size of the question sets (size 0 through size 77). When 100% accuracy was reached, the loop of testing was stopped, resulting in large sized question sets (starting around 43) to have fewer trials.

Each diagram in Fig. 16 represents one question set size ranging from 0 to 19 (size 20-77 are excluded from this paper due to page limitation). The diagrams show the accuracy distribution of the 1250 random sets for the given question set size.

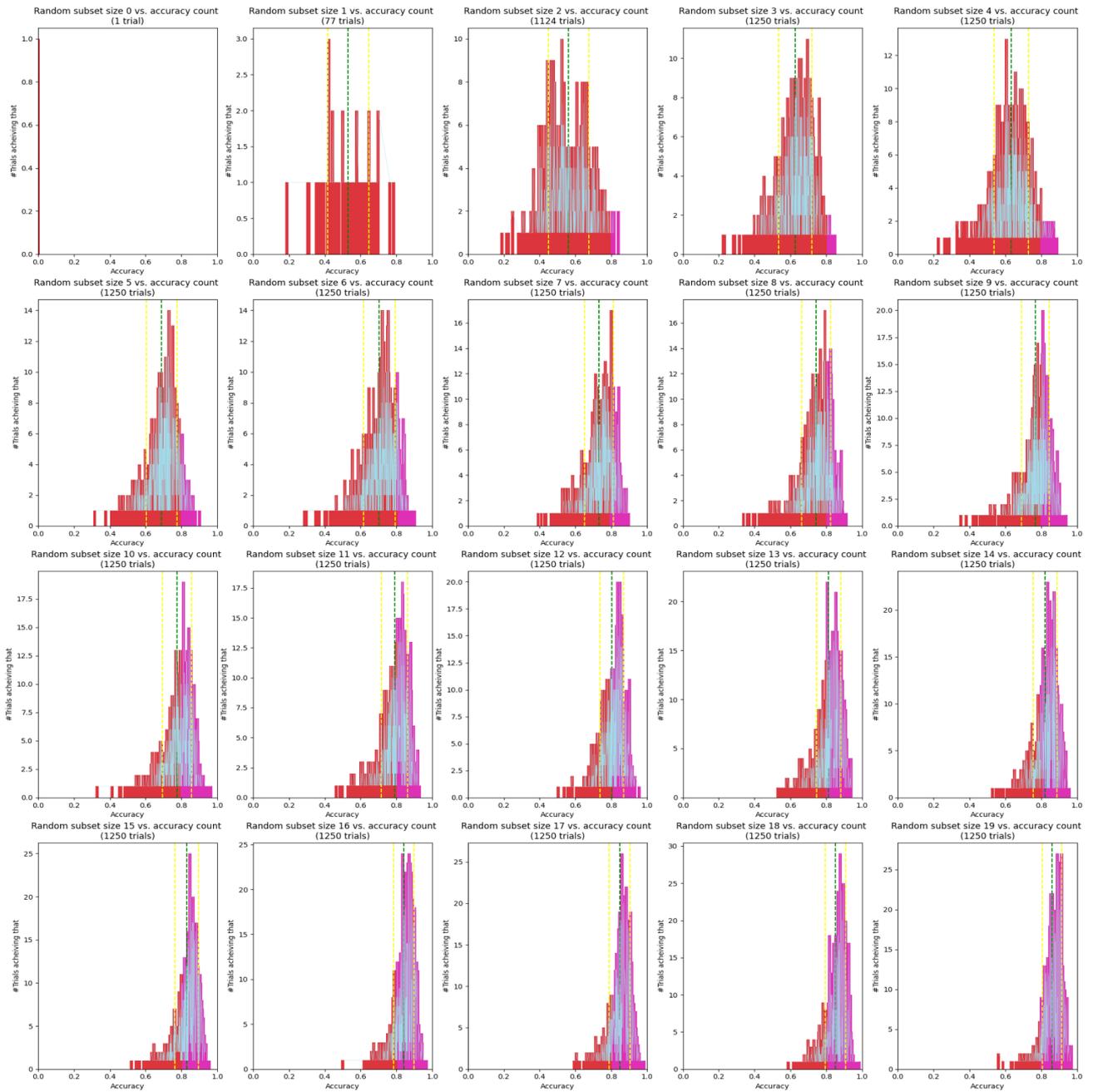

**Fig. 16**. Accuracy distribution of the 1250 randomly sampled subsets. Each diagram represents one subset size. X-axis represents the accuracy in percentage. Y-axis is the number of subsets that achieved the given accuracy. Green vertical dotted line represents mean accuracy, Yellow represents one standard deviation above and below. The question size of 13 reaches accuracy above 80%.

Notice that the 0.8 mean accuracy is achieved at question subset size of 13 and above. Final shortlist selection is ['I.1', 'I.2', 'I.8', 'I.9', 'II.13', 'II.15', 'II.6', 'III.10', 'III.14', 'III.15', 'III.9', 'IV.19', 'IV.5']

### IV. Next Steps

While our results are promising, the model was trained on data from a single provider with a relatively small sample size (60 subjects) and a wide age span. This limits the robustness of our analysis due to the lack of broader validation and the potential for overfitting from exhaustive subset search. Future work may explore nonlinear models to enhance accuracy while maintaining interpretability.

We plan to expand the dataset from the current 60 entries to a much larger sample and collect demographic and therapy metadata for each entry to enable more granular, cohort-based analyses. Specifically, we will form age cohorts to control for variance related to age and developmental stage. Given the limited sample size, the effect of age was inconclusive in the current study. We will also collect gender data (currently missing) and primary language information to evaluate potential differences, as prior research suggests that non-English speaking individuals may show greater improvement than English-speaking ones [37]. Furthermore, with a substantially larger sample, we can perform more robust validation to confirm our model's reliability and reduce the risk of overfitting.

In the future, we may also test our ML models on ATEC data across different interventions to validate their ability to customize assessments for diverse user cohorts. This approach can also be extended to diagnostic data from CARS-2, ADOS-2, ADI-R, and other tools to further evaluate model performance.

## V. Conclusion

In this research, we analyzed ATEC questionnaire data from 60 subjects undergoing applied behavior analysis (ABA) therapy, aiming to reduce the assessment burden while preserving diagnostic accuracy.

The proposed framework automatically estimated therapy effectiveness and provided actionable insights for the next steps in intervention adjustment. Our approach systematically identified the questions most predictive of the overall ATEC scores. Using multiple linear regression and cross-validation, we tailored the original 77-item questionnaire into a concise 16-question version while retaining coverage across all 4 ATEC subtests. This shortened form could potentially enable more frequent assessment and offer timely insights into therapeutic progress. Furthermore, we extended the framework to support Point-in-Time diagnosis, identifying a 13-question subset that moderately correlates with symptom severity, achieving over 80% accuracy.

As AI-based interventions are more widely applied in clinical settings, our research offers a pragmatic data-driven approach to streamline ASD assessment and enhance responsive care.